# Characterization of polarimetric based three axis atomic magnetometer

Swarupananda Pradhan[*] and Rita Behera

Laser and Plasma Technology Division, Bhabha Atomic Research Centre, Mumbai 400085, India

and

Homi Bhabha National Institute, Department of Atomic Energy, Mumbai-400094, India

[*]Corresponding author: spradhan@barc.gov.in, pradhans75@gmail.com

**Date: 31-07-2018**

**Abstract**

The operation of polarimetric based single beam three axis atomic magnetometer is studied. The magnetometer operates in presence of small bias magnetic field. Its salient features are compared with the analysis of Bloch equation. The sensitivity of the device along three orthogonal directions are measured in low frequency regime, where 1/f noise adversely affect the sensitivity. The cross talk among the three axes magnetic field measurements and relative phase shifts are analysed using Lissajous plot. The sensitivity of the magnetometer is found to be < 10 pT Hz$^{-1/2}$ @400 mHz in any direction. The investigated magnetometer is suitable for space application, where overall size of the device is of prime importance.

## I. Introduction:

The highly sensitive atomic magnetometers have promising prospect due to possible adaptability to variety of application-oriented conditions [1-5]. The achievable sensitivity has surpassed the conventional magnetometers while operating near room temperature [5-8]. Pioneering work of several research group has led to miniaturization of the sensor head that is a pre-requisite for bio-medical application [4, 7, 8]. The reduction in overall size and weight of these devices is an important parameter to position them as a viable replacement for the low sensitive flux gate magnetometers in satellites. For effective space utilization, there is a demand for compact and lightweight three axis atomic magnetometer with pT sensitivity in low frequency regime.

The three-axis atomic magnetometer for a variety of experimental configuration has been demonstrated [9-12]. Here we have extended the study on single beam three axis atomic magnetometer described in ref-10, that has important features for space applications. In this geometry, the magnetometry is carried out near atomic resonance and hence has additional option of *in-situ* laser frequency stabilization. The experiment is carried out at a lower temperature of the vapour cell that is close to the operating temperature of the diode laser, thereby easing thermal management of the device. The dynamic range of the device can be increased without compromising on the sensitivity by operating in close loop [13]. The operation with small magnetic coils around the atomic cell limits the homogeneity of the magnetic field in the probing volume. Since the magnetometer works in presence of bias magnetic field, it can overcome such practical limitation. In this article, we have investigated sensitivity of this magnetometer along three orthogonal directions in presence of bias magnetic field. The cross talk among three axis measurements are studied. A brief description of the related model based on Bloch equation and its application to the current experimental geometry is presented.

In the envisaged magnetometry geometry [10], an elliptically polarized light field along z axis perform the dual job of spin polarization through optical pumping and probing of the spin component along it. A small magnetic field is applied along y direction. Under this condition, oscillating magnetic fields are applied in three orthogonal direction to induce coherent oscillation of the atomic spin component along the probing direction [14, 15]. The amplitude of these coherent oscillation depends on the component of the ambient magnetic field along the respective oscillating field and are phase sensitively detected. The associated mechanism can be visualized in the following theoretical description.

## II. Theoretical description:

The interaction of a resonant light field with an atomic ensemble gets influenced by small changes in the orientation as well as amplitude of the magnetic field. The high sensitive measurement of such subtle changes is the basis of atomic magnetometry. The semi-classical density matrix calculation is a vital tool for detail analysis of the system. However, many important attributes of the involved mechanism can be addressed by the Bloch equation that describes the evolution of the spin polarization close to the experimental conditions [9, 11, 12, 14]. Since we are working near zero magnetic field, the dynamics of the spin can be approximated by the Bloch equation. The spin loss due to repolarization (due to laser beam) and diffusion can be neglected as a single elliptically polarized light beam is used to interact with atomic gas in a high-pressure buffer gas environment. Under these circumstances, the spin polarization $\boldsymbol{P}$ in a magnetic field $\boldsymbol{B}$ under steady state condition can be described by the Bloch equation as [14, 15]

$$\boldsymbol{P} \times \boldsymbol{B} + P_0 \Delta B = \Delta B\, \boldsymbol{P}$$

Where $P_0 = s\, R/(R + \Gamma_{Re})$, $\Delta B = (R + \Gamma_{Re})/\gamma$, $s$ is the equilibrium spin polarization along the laser propagation direction, $R$ is the optical pumping rate, $\Gamma_{Re}$ is the spin relaxation rate, and $\gamma$ is the gyromagnetic ratio. The observed narrow resonance width (for the kink structure ~200nT, ref-10) in our experimental conditions, indicates the onset of diminishing role of spin exchange relaxation mechanism.

The optical pumping and probing of spin polarization is carried out by a single elliptically polarized light along the z axis. The reflected light field across the detection PBS is used for the measurement of the magnetic field. Thus, the observed signal is related to the optical rotation of laser beam due to spin polarization along z axis ($P_z$). Unlike prior-works with neglected bias field [9, 11, 12, 14], we are interested in studying the response of $P_z$ to three axis magnetic field in presence of small bias-fields. Thus, the steady state solution of Bloch equation without neglecting any contribution to $P_z$ becomes

$$P_z = P_0 \frac{B_z(B_x + B_y + B_z) + \Delta B(B_y - B_x) + \Delta B^2}{B_x^2 + B_y^2 + B_z^2 + \Delta B^2}.$$

The $B_x$, $B_y$ and $B_z$ magnetic fields are imposed with modulations at frequency at $\omega_x$, $\omega_y$, and $\omega_z$ respectively. The demodulated light intensity (after the cell) at $\omega_x$, $\omega_y$, and $\omega_z$ will corresponds to $\frac{\partial P_z}{\partial B_x}$, $\frac{\partial P_z}{\partial B_y}$, and $\frac{\partial P_z}{\partial B_z}$ respectively. It can be found that for $B_z = 0$, all the demodulated signal profiles will cross zero simultaneously for $B_x = -(\sqrt{3} - 1)\Delta B/2$ and $B_y = +(\sqrt{3} - 1)\Delta B/2$ as shown in Fig-1. The $\frac{\partial P_z}{\partial B_x}$ is very sensitive to small change in $B_x$ field and insensitive to the $B_y$ field around the respective bias field. Thus $\frac{\partial P_z}{\partial B_x}$ can be used for measurement of $B_x$ field. Similarly, $\frac{\partial P_z}{\partial B_y}$ can be used for measurement of the $B_y$ field. However, $\frac{\partial P_z}{\partial B_x}$ and $\frac{\partial P_z}{\partial B_y}$ also depends on the $B_z$ field though with an opposite polarity (with respect to $B_x$ and $B_y$ field respectively). Similarly, the response of $\frac{\partial P_z}{\partial B_z}$ on $B_z$ field is opposite to that with $B_x$ and $B_y$ field. These dependencies can be judiciously used for simultaneous measurement of three axis magnetic field.

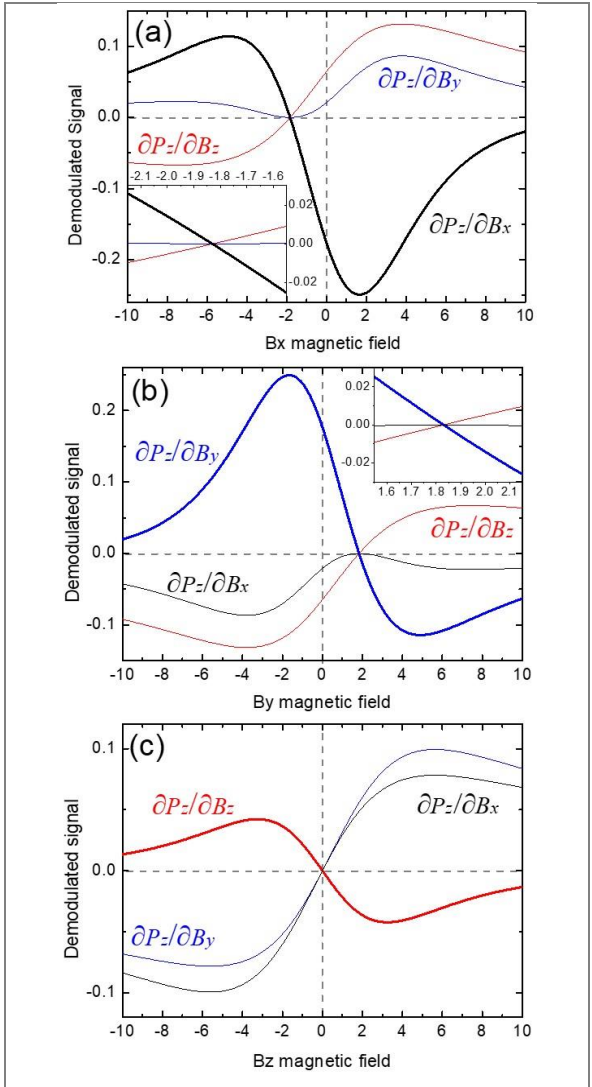

**Fig.-1:** The calculated changes in $P_z$ with respect to change in $B_x$, $B_y$, and $B_z$ field are shown for scanning of magnetic field along three directions. For scanning in any direction, the orthogonal bias fields are kept at $B_x = -(\sqrt{3}-1)\Delta B/2$, $B_y = +(\sqrt{3}-1)\Delta B/2$, $B_z = 0$, and $\Delta B = 5$.

The above analysis is carried out for a special case of $B_z = 0$. For any finite value of $B_z$ field, there exist a set of orthogonal fields $B_x = (\sqrt{3}-1)(B_z - \Delta B)/2$ and $B_y = (\sqrt{3}-1)(B_z + \Delta B)/2$ around which three axis magnetometry can be performed. This indicates interdependency among bias fields and are not unique. However, experimentally we found a unique set of bias fields for $B_x$, $B_y$ and $B_z$ where all the demodulated signal goes to zero [10, 16, 17]. Further, this simple model overestimates the experimentally observed cross talk and is discussed in Sec-IV. In practice, the signal profile is governed by the competing role of optical pumping followed by Zeeman redistribution and quantum interference that are function of experimental conditions like buffer gas pressure, polarization state of the input light, tilt between the magnetic field and laser propagation directions, and others [17-21]. A more elaborate density matrix calculation can address these subtle issues pertaining to the phenomena and is in progress. Despite its limitation, the analysis of Bloch equation is useful to visualize the underlying mechanism.

### III. Experimental apparatus:

The experiment is carried out with a single elliptically polarized laser beam as shown in Fig-2 [10, 16]. The temperature of the atomic cell (natural Rb atoms at 25 torr $N_2$ gas) is kept at $48^0$ C. The magnetic field at the probing volume is controlled by few layers of mu-metal sheets and three set of coils placed around the cell. The magnetic coils are calibrated with respect to coherent population trapping signal [10, 22]. The transmitted and reflected light across the detection PBS are used for laser frequency stabilization and magnetic field measurement respectively.

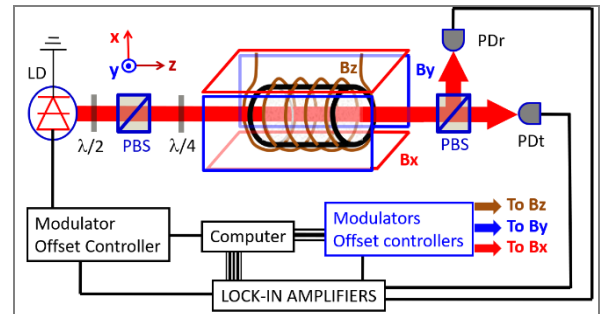

**Fig-2:** The magnetic field is measured by demodulating the reflected photodiode signal (PDr) with respect to modulation applied in three directions.

The injection current of the LD (laser diode) is modulated at 12 kHz to impose a modulation in the laser frequency. The transmitted light across the PBS is phase sensitively detected with respect to the applied modulation for generating error signal for frequency stabilization. The laser frequency is stabilized in between the $^{85}$Rb atomic transition using the demodulated signal [10, 13]. The magnetic field along x, y and z direction are modulated at 69 Hz, 79 Hz and 55 Hz respectively and the corresponding phase sensitively detected signal acquired with analog lock-in amplifier (Scitec-420) are named $MM_xR$, $MM_yR$ and $MM_zR$ respectively ($MMR$ represents all of them). These signals are calibrated with respect to the current through the coils and correspond to $\frac{\partial P_z}{\partial B_x}$, $\frac{\partial P_z}{\partial B_y}$, and $\frac{\partial P_z}{\partial B_z}$ respectively. The amplitude of modulation along x and y axis is 175 nT, where as it is 85 nT for z axis. These are optimized values against amplitude, width and noise in the corresponding signal. The $MM_xR$, $MM_yR$ and $MM_zR$ signal are iteratively bought close to zero by changing corresponding bias field [10]. This establishes the operating condition of the magnetometer where after magnetic field components are measured from the corresponding $MMR$ signal. In a close loop operation, bias current through the coils along with corresponding $MMR$ signal provides the component of the ambient magnetic field [13].

## IV. Results and discussions:

The operating condition of the magnetometer is established by iteratively changing the bias magnetic fields to bring the $MM_xR$, $MM_yR$ and $MM_zR$ signal close to zero. This occurs for bias fields of ~-4nT, +8.5 nT and +128 nT along x, y and z direction respectively for quarter wave plate at $20^0$. The observed unique values of bias field don't agree well with the analysis of Bloch equation. The signals under this condition are shown in the initial ~30 seconds of the Fig-3, where drift in the signals represent corresponding change in the component of the ambient magnetic field. The $B_x$ magnetic field is changed by ±1nT in the time period of ~30 to ~80 second. It shows that the change in $B_x$ field is reflected in the $MM_xR$ signal. Further, it doesn't show any cross talk in $MM_yR$ signal, consistent with the inset of Fig-1(a). However, the predicted cross talk of $B_x$ field on $MM_zR$ signal is not observed. Similarly, only $MM_yR$ signal get affected by the change in $B_y$ magnetic field. The change in the $B_z$ field is reflected in the $MM_zR$ signal as well as in the $MM_yR$ signal, albeit the later with a smaller amplitude. The opposite polarity of the cross talk in $MM_yR$ signal agrees well with the result of Bloch equation shown in the Fig-1 (c), though observed amplitude is overestimated. Also, the predicted dependence of the $MM_xR$ signal on the $B_z$ field is not observed. As has been pointed out, a more elaborate density matrix calculation is required to capture the experimental observations. The observations indicate that $B_x$ and $B_z$ magnetic field can be measured from the $MM_xR$ and $MM_zR$ signal respectively. The $B_y$ magnetic field can be measured from the $MM_zR$ signal by correcting the cross talk due to $B_z$ magnetic field appropriately.

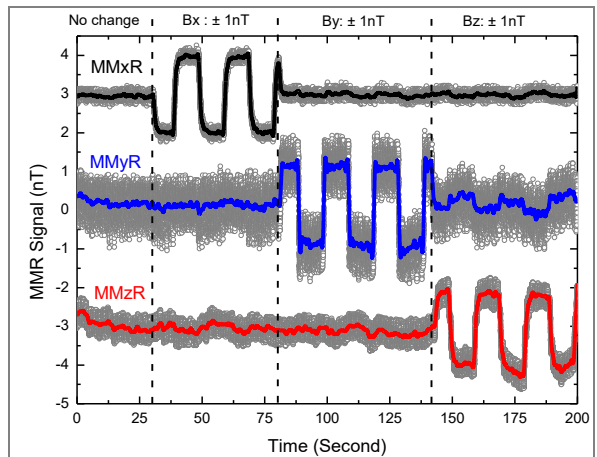

**Fig-3:** The $MMR$ signals are shown for sequential change in DC magnetic field by ±1nT along x, y, and z direction. The $MM_xR$ and the $MM_zR$ signal are shifted by +3 nT and -3nT respectively for better visualization. The hollow circles are experimental data and the solid lines are post acquisition processed data with a 2Hz low pass FFT filter.

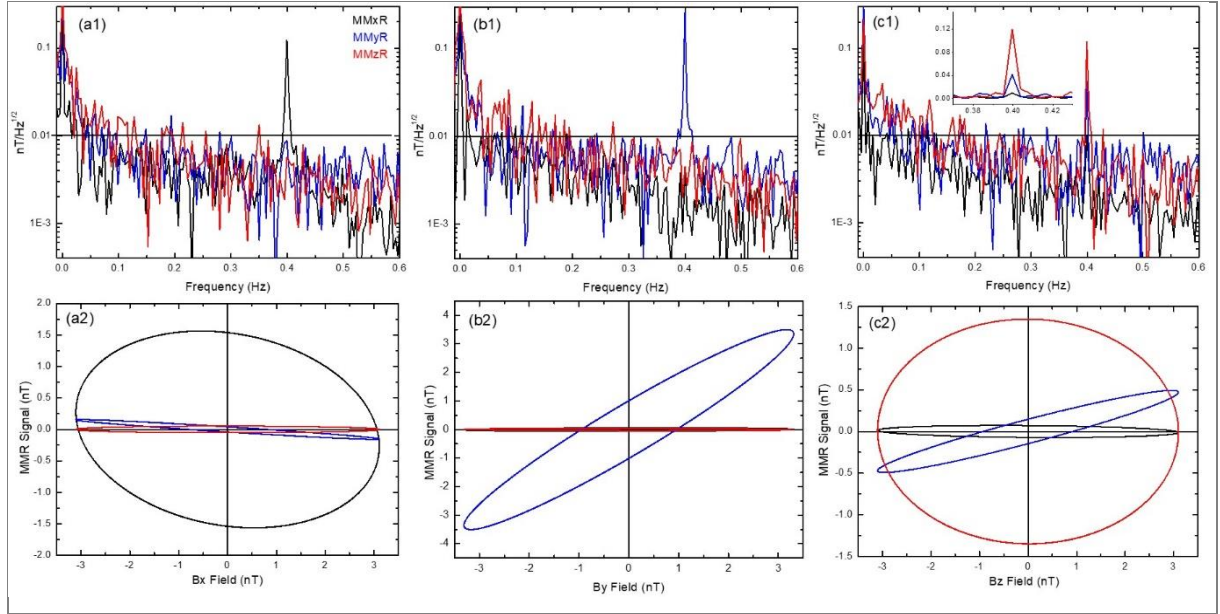

**Fig-4**: The magnetic field noise spectral densities are measured simultaneously along three orthogonal direction. The black, blue and red curve represents the $MM_xR$, $MM_yR$ and $MM_zR$ signal respectively. The noise densities are primarily limited by the electronic noise. For a1, b1 and c1, a magnetic field modulation at 400 mHz with an amplitude of 300 pT is applied along x, y and z direction respectively. The corresponding Lissajous plot shown in a2, b2, and c2 are obtained by modulating the magnetic field at 400 mHz with an amplitude of ±3nT.

The spectral magnetic noise of the component of magnetic field in three axes while applying a small magnetic field @400 mHz in one of the axis are shown in Fig-4(a1, b1, c1). The sensitivity of the magnetometer is < 10 pT Hz$^{-1/2}$ in any of the direction for the investigated frequency range. The spectral response of the $MMR$ signals with $B_x$, $B_y$, and $B_z$ fields are consistent with the observation of Fig-3. Similar to response of $MMR$ signal to change in components of DC magnetic field, the Bloch equation doesn't explain the observed cross talk for small changes in magnetic field (amplitude ~300 pT @400mHz).

The crosstalk among measurements of magnetic field components and polarity of the signals predicted in Fig-1 is further analysed by applying relatively larger change in the magnetic field. The Lissajous plot is an ideal tool to study the cross talk, phase relationship and transfer function of the magnetic field measurement among the three axes. This is realized by applying ~±3nT @ 400mHz along one axis and taking the time series data of $MM_xR$, $MM_yR$ and $MM_zR$ simultaneously. In contrary to observation related to small change in magnetic fields, the $MM_xR$ and $MM_yR$ signal are also influenced by change in $B_y$ and $B_x$ field respectively, apart from predicted cross talk in Fig-1. Here also, the amplitude of the cross talk doesn't agree with the results of the Bloch equation. The measured magnetic field is half of the applied magnetic field. This is due to attenuation by the home made analog roll-off filter that is used after the lock-in amplifier. Though the $MM_yR$ signal doesn't show any attenuation, the roll off filters used for $MM_xR$ and $MM_zR$ signal are attenuated by half at 400 mHz. The implementation of digital filters and better quality of lock-in amplifier can avoid such problem.

The extraction of the phase of the signals with respect to the applied oscillation is complicated due to contribution of the inherent phase shifts occurring at different electronics components and magnetic coils.

Further, the electronics components used for signal processing of $MM_xR$, $MM_yR$ and $MM_zR$ are slightly different and also the magnetic coils used for generating $B_x$, $B_y$, and $B_z$ magnetic fields. Nevertheless, the comparison of Lissajous plot shown in Fig-4 a2, b2 and c2 can be useful to obtain the relative phase shift of any $MMR$ signal with respect to change in magnetic field in three orthogonal directions. The magnitude of phase shift from the Lissajous plot is given by $\Delta\theta = \pm \sin^{-1}(\Delta x_0/\Delta x_m)$ for first quadrant and $\Delta\theta = \pm[180^0 - \sin^{-1}(\Delta x_0/\Delta x_m)]$ for second quadrant, where $\Delta x_0$ and $\Delta x_m$ are the zero crossing horizontal width and maximum horizontal width of the ellipse respectively. The sensitivity of the signal to change in the field is related to $\Delta y_m/\Delta x_m$, where $\Delta y_m$ is the maximum vertical width. We have utilized Fig-1 to fix the ambiguity of $\pm$ sign in the phase shift and assume that electronics phase shift is fixed for a $MMR$ signal with respect to any change in the magnetic field components. The resultant change in the phase shift is given in the table-1. The observed phase shift of $MM_xR$ & $MM_yR$ signal with $B_z$ field, and $MM_zR$ signal with $B_x$ & $B_y$ field are consistent with Fig-1.

|       | $MM_xR$ | $MM_yR$ | $MM_zR$ |
|-------|---------|---------|---------|
| $B_x$ | 0       | 182     | 180     |
| $B_y$ | 176     | 0       | 185     |
| $B_z$ | 201     | 180     | 0       |

**Table-1:** Relative phase shift of the $MMR$ signal for change in magnetic field components.

It may be possible to extract important information on absolute phase shift of $MMR$ signal by having a control over the electronics phase shift. The observed phase shift provides another handle to discriminate the crosstalk from the measurement. The reflected light intensity across the detection PBS demodulated at the modulation frequency of the laser current [13] can provide an auxiliary information on the $B_z$ field (not utilized in this work).

## V. Conclusions:

The magnetometer responses to the change in magnetic field along three axes are analysed and compared with the outcome of the Bloch equation. The sensitivity of the magnetometer along three axes are simultaneously measured. The magnetometer operates with unique set of bias magnetic field along three directions, and laser frequency is *in situ* stabilized. The overall device can be made compact as required for space application. The cross talk of measurement with respect to change in magnetic field in orthogonal directions are presented. The observed phase shift can be another handle to discriminate the cross talk from the measurement.

**Acknowledgements:**

The authors are thankful to Mr. R. K. Rajawat, Asso. Dir. BTDG for supporting the activity.